\documentclass[
]{ceurart}

\usepackage{listings}
\usepackage{amsthm}
\usepackage{float}
\usepackage{url}
\usepackage{booktabs}
\usepackage{listings}
\usepackage{color}
\usepackage{xcolor}
\definecolor{mygrey}{gray}{0.9}
\definecolor{mygreen}{RGB}{208,240,192}

\begin{document}

\copyrightyear{2022}
\copyrightclause{Copyright for this paper by its authors.
  Use permitted under Creative Commons License Attribution 4.0
  International (CC BY 4.0).}

\conference{CLEF 2022: Conference and Labs of the Evaluation Forum, 
    September 5--8, 2022, Bologna, Italy}

\title{Few-shot Long-Tailed Bird Audio Recognition}

\author[1, 2]{Marcos V. Conde}[%
email=marcos.conde-osorio@uni-wuerzburg.de,
url=https://mv-lab.github.io/,
]

\fnmark[1]
\address[1]{H2O.ai}
\address[2]{Computer Vision Lab, Institute of Computer Science, University of Würzburg, \sep Germany}

\author[3]{Ui-Jin Choi}[%
email=choiuijin1125@megastudy.net,
]
\fnmark[1]
\cormark[1]
\address[3]{MegaStudyEdu, \sep South Korea}

\cortext[1]{Corresponding author.}
\fntext[1]{Authors contributed equally.}

\begin{abstract}
It is easier to hear birds than see them. However, they still play an essential role in nature and are excellent indicators of deteriorating environmental quality and pollution. Recent advances in Deep Neural Networks allow us to process audio data to detect and classify birds. This technology can assist researchers in monitoring bird populations and biodiversity.
We propose a sound detection and classification pipeline to analyze complex soundscape recordings and identify birdcalls in the background. Our method learns from weak labels and few data and acoustically recognizes the bird species. Our solution achieved 18th place of 807 teams at the BirdCLEF 2022 Challenge hosted on Kaggle.\\
Code and models will be open-sourced at \url{https://github.com/Choiuijin1125/bclef2022}.
\end{abstract}


\maketitle

\section{Introduction}
\label{sec:intro}

The BirdCLEF 2022 Challenge~\cite{kahl2022overview, lifeclef2022} proposes to identify which birds are calling in long recordings given quite limited training data. This is the exact challenge faced by scientists trying to monitor rare birds in Hawaii. However, we propose a novel machine learning solution to help advance the science of bioacoustics and support ongoing research to protect endangered Hawaiian birds. 

The \textbf{motivation} behind this challenge and our solution is the fact that Hawaii has lost 68\% of its bird species~\cite{challenge}.

Researchers use population bioacoustic monitoring to understand how native birds react to changes in the environment and conservation efforts. This approach could provide passive, low labor, and cost-effective strategy for studying endangered bird populations. Current methods for processing large bioacoustic datasets involve manual annotation of each recording. This is an expensive process that requires specialized training and large amounts of time. For this reason, we propose a Machine Learning solution to automatically identify bird species in long audio recordings via birdcall detection and classification within the audio.

\begin{figure}[h!]
    \centering
    \setlength{\tabcolsep}{2.0pt}
    \begin{tabular}{cccc}
    \includegraphics[width=0.25\textwidth]{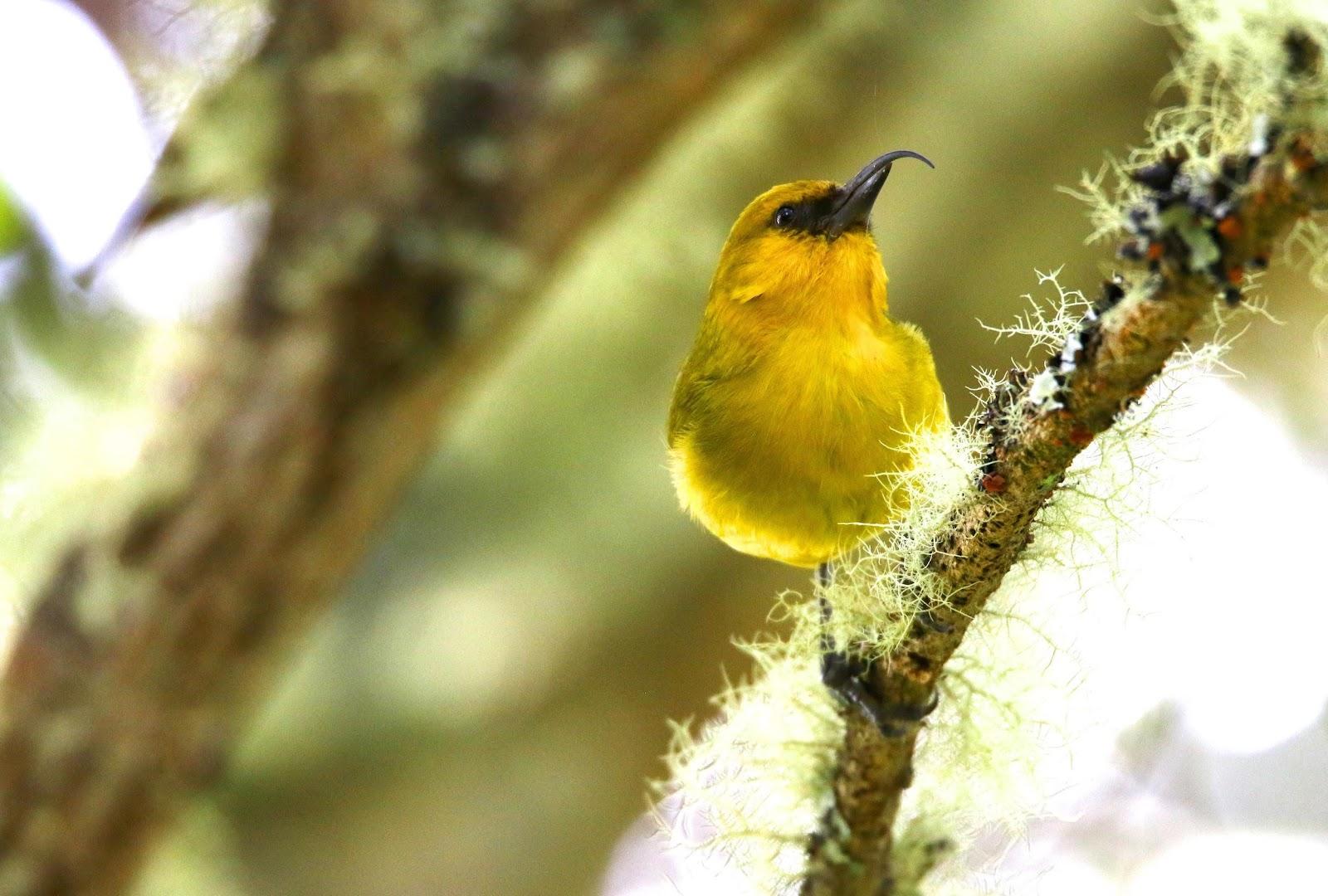} & 
    \includegraphics[width=0.25\textwidth]{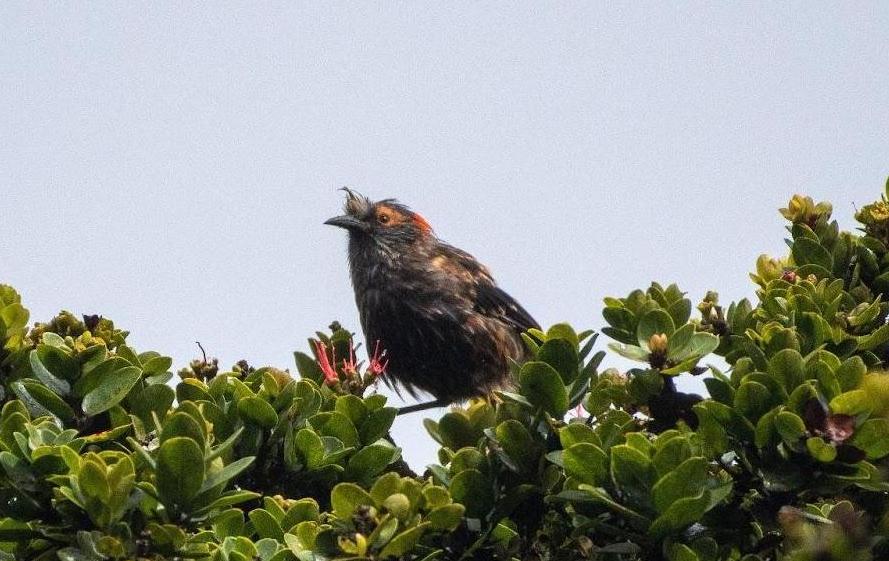} &
    \includegraphics[width=0.25\textwidth]{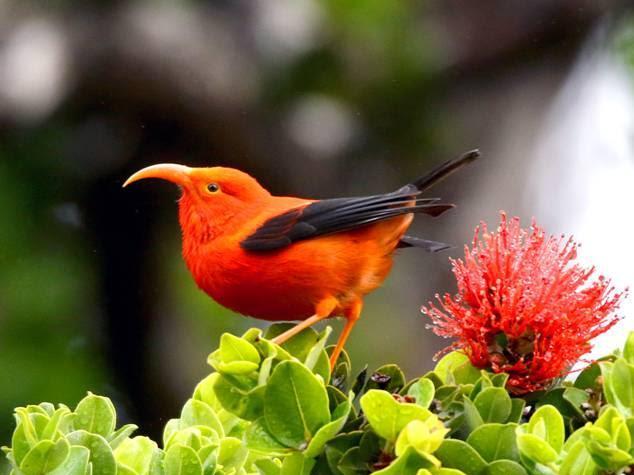} &  \includegraphics[width=0.25\textwidth]{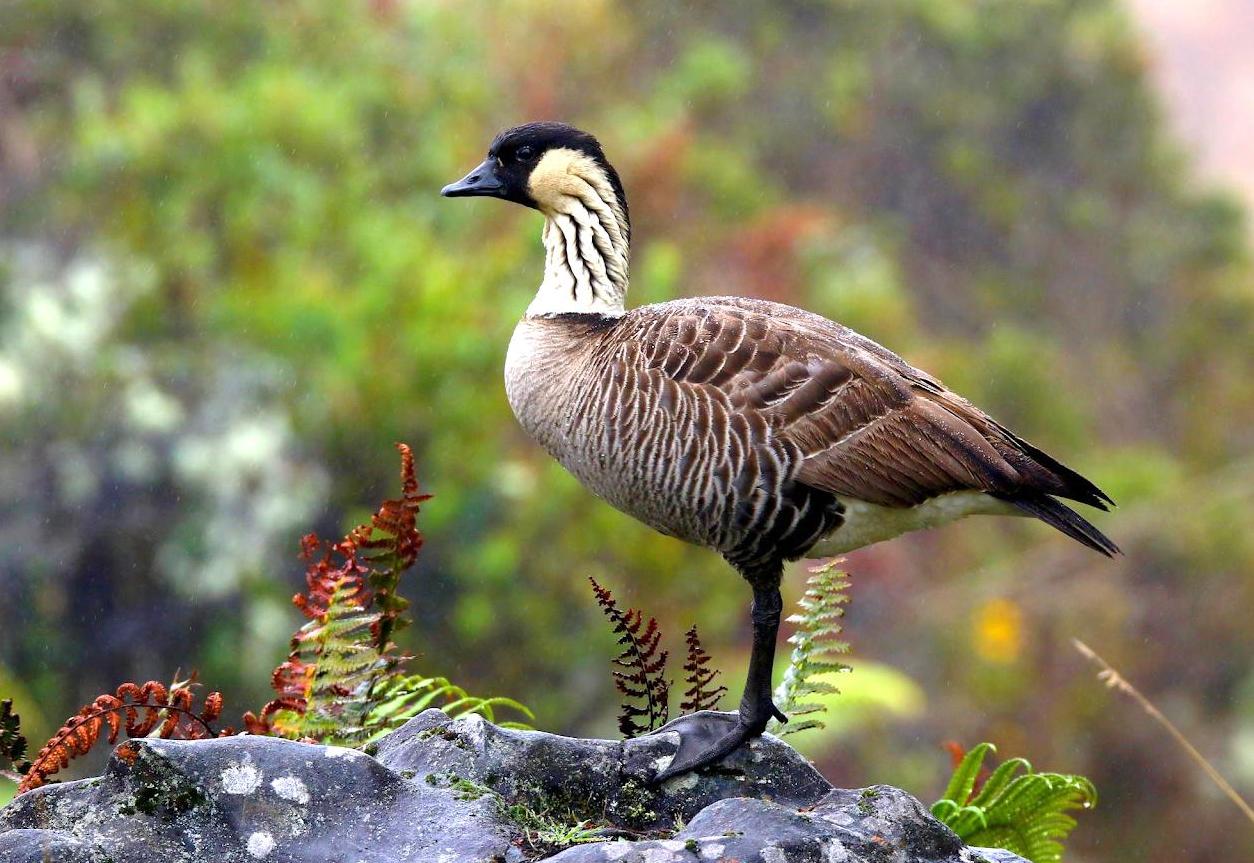}\tabularnewline
    'Akiapōlā'au & 'Ākohekohe & 'I'iwi & Nēnē \tabularnewline
    Hemignathus wilsoni & Palmeria dolei & Drepanis coccinea & Branta sandvicensis \tabularnewline
    \end{tabular}
    \caption{Photographs of some Hawai'i endemic bird species studied in this work. Photo credit: Amanda K. Navine, Alexander Wang and Ann Tanimoto-Johnson.}
    \label{fig:birds}
\end{figure}

\subsection{Related Work}

Recent advances in Machine Learning (ML) have made it possible to automatically identify bird songs for common species using annotated soundscapes as training data. 

The main challenges from the machine learning point of view are:
\begin{enumerate}
    \item Weak labels. Training data consists of soundscapes of variable duration, recorded in the wild. Therefore, we find substantial noise in the recordings (other birds besides the target, rain, wind, planes, etc).

    \item Long-tailed distribution. Rare and endangered species (such as those in Hawaii) are less represented in the training data, and therefore, the model struggles to learn their features and generalize for those classes.
    
    \item Few-shot training is required. We find less than four recordings for some endemic bird species (crehon, hawhaw, maupar, etc). The most represented bird is ''skylar" with 500 recordings, which is not a tremendous amount of training data in the context of ML.

\end{enumerate}

Previous years BirdCLEF challenges~\cite{birdclef2021, inproceedings_birds2020} proposed different problems related to large-scale bird recognition in soundscapes or complex acoustic environments.
Sprengel~\textit{et.al.} \cite{Sprengel2016AudioBB} and Lasseck \cite{Lasseck2019BirdSI,Lasseck2018AudiobasedBS} introduced deep learning techniques for the ''Bird species identification in soundscapes" problem.
State-of-the-art (SOTA) solutions are based on Deep Convolutional Neural Networks (CNNs) \cite{Schlter2018BirdIF, Bai2020XceptionBM_birds, Mhling2020BirdSR}, usually, deep CNNs with attention mechanisms are selected as backbone in these experiments \cite{zhang2020resnest, tan2020efficientnet, kong2020panns, he2015deep}, or suitable for fine-grained classification tasks~\cite{conde2021exploring}.
Pretrained audio neural networks (PANNs) \cite{kong2020panns} provide a multi-task SOTA baseline for audio related tasks, showing great generalization capability.
Other approaches are focused on Sound Event Detection (SED) \cite{inproceedings_sed2020, fonseca2019learningsed, kong2020panns, article_sed}, similar to video understanding~\cite{zhang2019multi}, these approaches usually employ 2D CNNs to extract useful features from the input audio signal (log-melspectrogram), these features still contain information about frequency and time, then recurrent neural networks (RNNs) are used to model longer temporal context from the extracted features or use the feature map directly to predict since it preserves time segment information.

Solutions for the BirdCLEF 2021 Challenge follow these directions, moreover, they propose additional post-processing techniques to eliminate false detections (FPs)~\cite{murakami}, divers CNN-based ensembles~\cite{Henkel, Marcos, das2021solution, Kumar2021solution, Shugaev2021solution, jan2021}, and transformer-based solutions like STFT~\cite{stft2021}. These solutions can identify birds in long audio recordings, at different locations (Colombia, USA, and Costa Rica), with $\approx$ 70\% accuracy. In this challenge, we focus only on Hawaiian Bird Species.

We define some terms related to this challenge referring last year competition solution~\cite{Marcos, Henkel} that we will use in the description of our method in Section~\ref{sec:method}:

\begin{itemize}
\item Leaderboard denoted as LB (including its two variants, public and private)

\item Cross-Validation denoted as CV.

\item We define ''nocall" as the class corresponding to the events (a.k.a segments or clips) in audio where birdcalls are not detected.

\item We refer to the ''BirdCLEF 2021 Birdcall Identification Challenge (Kaggle)" as the "previous or last competition".

\item We define ''weakly labeled" as the labels that do not contain time-wise information about bird species in audio clips (i.e., not specific information about in which 5s segment in the audio, the bird calls).

\item We define ''strongly labeled" as the labels that contain time-wise information about bird species in audio clips (i.e., approximate second within the audio, where the bird calls).

\item BirdCLEF 2021 train soundscapes audios denoted as ''train soundscapes" which are 20 audio clips strongly labeled.
\end{itemize}

\subsection{Dataset}
\begin{figure}[!ht]
  \centering
  \includegraphics[width=\linewidth]{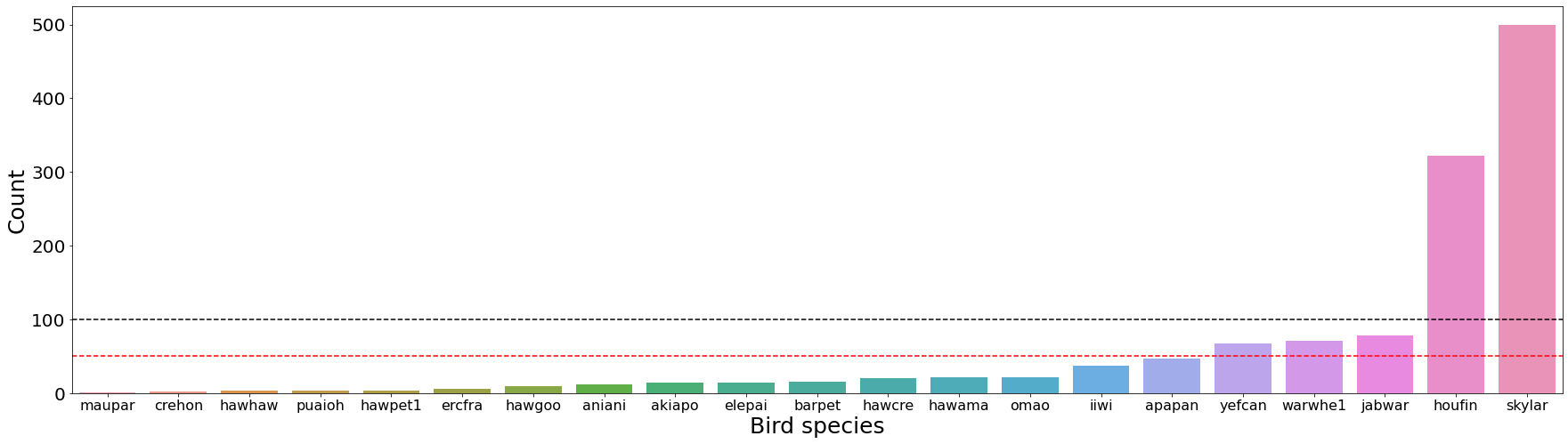}
  \caption{Distribution of bird species in the training set. We can see a notable long-tailed distribution. Many bird species are represented with less than 10 audio clips (i.e., maupar, crehon, hawhaw, puaioh). The red line indicates that most of the birds appear less than 50 times in the training set. Only two birds (houfin and skylar) appear more than 100 times in the training data.}
  \label{figure1}
\end{figure}

The training set consists of short audio recordings of 152 bird species, and only 21 bird species of interest are scored. These bird species inhabit Hawaii. However, many of the remaining birds across the islands are isolated in difficult-to-access, high-elevation habitats. Therefore, physical monitoring is difficult, and scientists have turned to sound recordings. As we show in Figure \ref{figure1}, the distribution of the ''interesting" bird species is very long-tailed\cite{GLD}, making it necessary to deal with extreme class imbalance. As we introduced, in this competition, our challenge is to develop ML models to identify bird species using sounds. Such models have to deal with real-world problems such as long-tailed rare birds and weak-noisy labels.

\subsection{Evaluation}
The performance is measured using a custom metric that is most similar to the ``macro F1 score". The test set consists of approximately 5500 recordings. Participants submit code and models and never have access to the test audios. There is a public LB that shows the score corresponding to 16\% of the test (880 audios), and a private or final LB with the scores over the rest 84\%.

\section{Methods}
\label{sec:method}

\subsection{Preprocessing}
Previous BirdCLEF challenges~\cite{birdclef2021, inproceedings_birds2020} showed that long audio clips for training improve performance. For this reason, we randomly cropped a 30-second time window of each audio, next, we split the 30s audio clips into 5s and 6-parts chucks as proposed by Henkel~\textit{et al.}~\cite{Henkel}, finally we transformed the such chunks to Mel Spectrogram using torchaudio library. 
The spectrograms were generated using the following parameters: sample\_rate=32000, n\_mels=128, fmax=14000, fmin=50 hop\_size=512, hop\_size=512,top\_db=None. 

\subsection{Augmentations}
After splitting audios, we applied 3 types of augmentations to handle robustness and the long-tailed distribution problem. First, we used three external datasets \textbf{freefield1010}\cite{freefield1010}, \textbf{BirdVox-DCASE-20k}\cite{lostanlen_vincent_2018_1208080}, \textbf{train\_soundscapes} (from 2021 Challenge) for background noise.

Second, to handle class imbalance, we used selective mixup~\cite{Zhang} which only uses the 21 scored birds of interest in the audio clip. In every training batch, we fed randomly cropped scored birds Mel Spectrograms and used mixup with training data. Next, we applied spec-augmentations\cite{Park_2019}. This method showed good performance in our local validation (CV), especially selective mixup boosted + 0.03 our score. 

\begin{figure}[!ht]
  \centering
  \includegraphics[width=\linewidth]{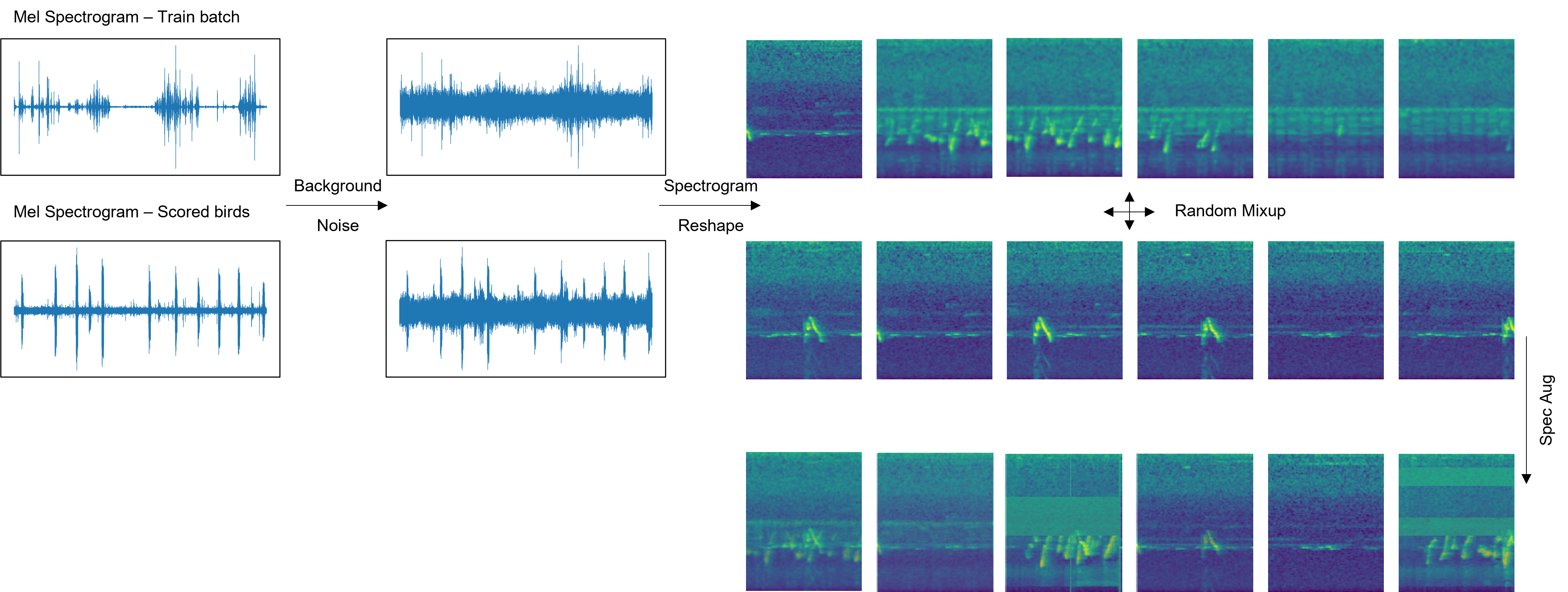}
  \caption{Illustration of our pre-processing pipeline. In every training batch, we used mixup of ''scored" birds. We can expect that selective mixup can make very robust training for rare birds~\cite{Henkel}.}
  \label{figure2}
\end{figure}

However, we also observed an overfitting behavior in some classes. because some scored birds (21 classes as in Figure~\ref{figure1}) have a long-tail distribution, model tends to predict more high confidence scores for some birds like a ``skylar" and ``houfin" which has a large distribution. Figure \ref{figure2} shows the pre-processing pipeline and augmentations.  

\subsection{Modeling}
\begin{figure}[!ht]
  \centering
  \includegraphics[width=\linewidth]{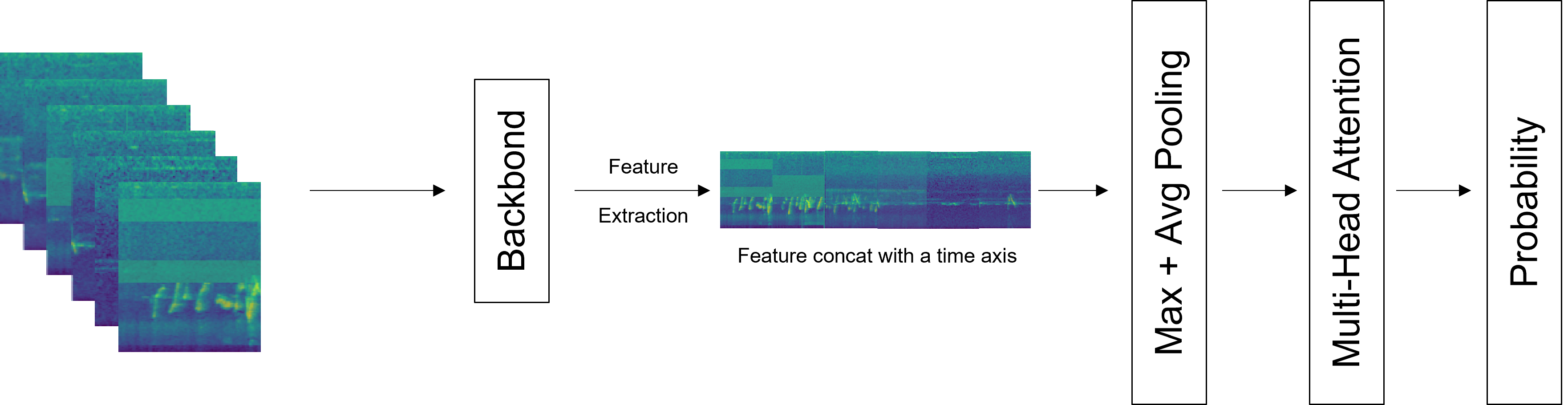}
  \caption{Illustration of our Bird Classification model inspired in~\cite{Henkel}.}
  \label{figure3}
\end{figure}

We used 9 different backbones and 22 models. Our models are a combination of top solutions from previous competition~\cite{birdclef2021}. The main architectures are CNNs backbones with Sound Event Detection heads~\cite{article_sed}, which showed good performance in the previous challenges~\cite{birdclef2021}. We feed 5-second, 6-part Mel Spectrograms into the network as~\cite{Henkel}. Figure \ref{figure3} shows the pipeline of our models. We also tried ConformerSED\cite{Miyazaki2020CONFORMERBASEDSE}, FDY-SED\cite{Hyeonuk}, HTS-AT\cite{HTSAT} but the results were much worse than using well-known CNN approaches. We show the different backbones in Table~\ref{tab:experiments}. In particular we focus on \texttt{tf\_efficientnet\_b0\_ns} since it is light-weight and suitable for smartphone devices, and its performance is consistently competitive.

\subsection{Training}
We trained our models using focal binary cross-entropy loss, AdamW optimizer, and cosine annealing scheduler with batch size 24. We also used the quality rating, which is meta information on audio quality. While computing the loss, we weight it using the normalized quality rating, and we used one-sided label smoothing, adding 0.01 across all negative labels. Both methods were proposed by Henkel~\textit{et al.}~\cite{Henkel} and improved our performance consistently.

\subsection{Post-Processing}

\subsubsection{Penalization}
We observed that our models are biased, and tend to predict with high confidence scores the most represented birds (i.e. skylar and houfin). This implies a large number of False Positives (FP) and misclassified clips. We give a penalty score depending on the distribution of the birds, such that most represented birds are more penalized. Penalization (PN) can be explained as in Eq.\ref{penalization} where \emph{x} is the distribution of each bird. We used \emph{penalty factor}=0.8. Penalization is not a realistic technique, the model should not filter out bird species in that way, yet, in this scenario it boosted our score on the public LB. As a better alternative, we aim to make our method less sensitive to the data distribution and robust against background birdcalls from non-interest birds, we tried to find class-wise thresholds of each bird species.

\begin{equation}
\label{penalization}
\textit{p}_{i} = \textit{p}_{i} - \textit{penalty factor}\times\frac{x_i}{\sum_i^nx_i}
\end{equation}

\subsubsection{Class-Wise Thresholds}
\label{sec:class-wise}

\begin{figure}[!ht]
  \centering
  \includegraphics[width=\linewidth]{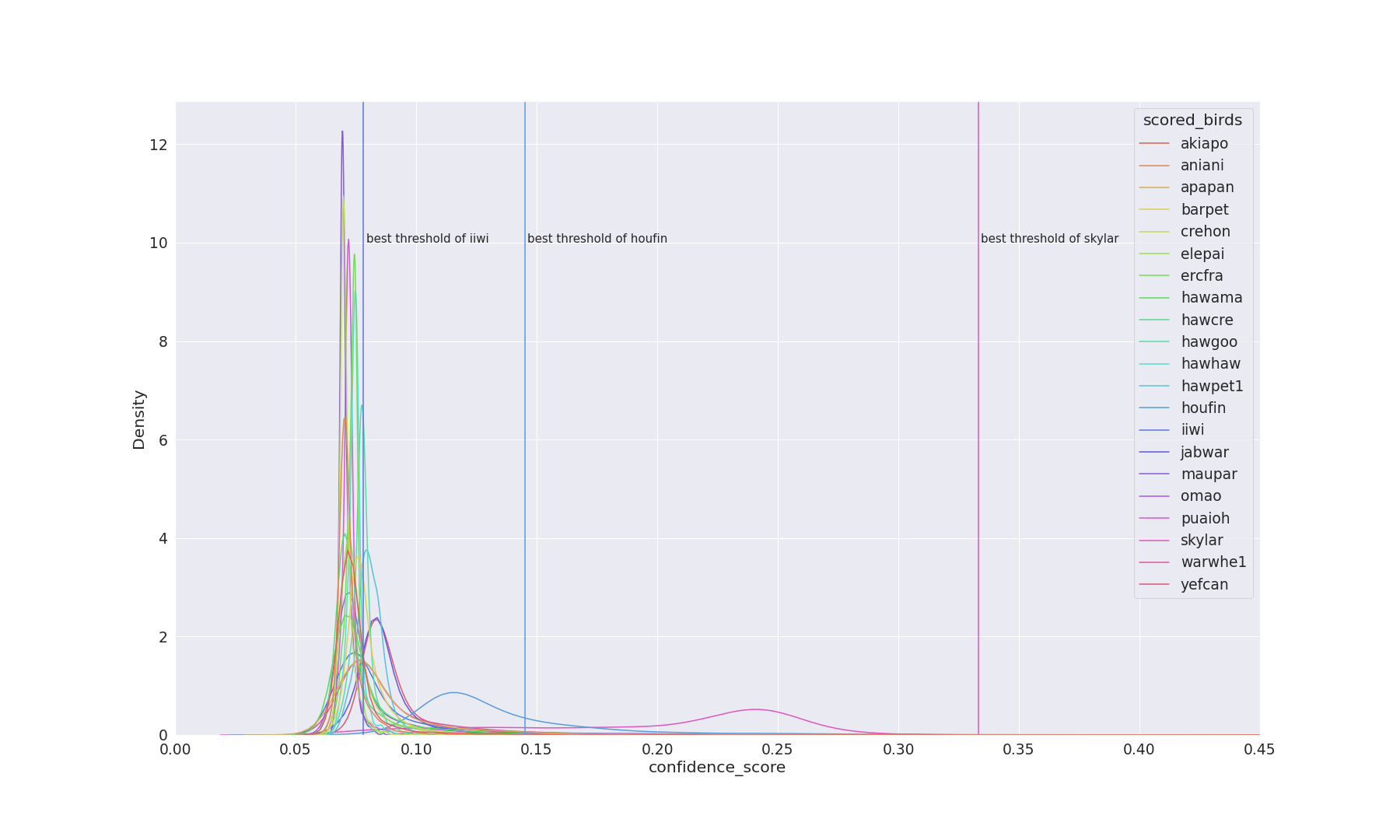}
  \caption{Distributions of nocall probability validated using train\_soundscapes. We show the class-wise best quantile thresholds to obtain the maximum AUC score per bird.}
\label{figure4}
\end{figure}

We observed that if there is a birdcall in the audio clips, regardless of the bird species, our models show higher confidence scores. We used train\_soundscape audio clips to validate nocall thresholds for bird species using AUC score per each bird (as a binary classification problem call/nocall). Even though there is no label for scored birds in train\_soundscape, we can estimate the appropriate nocall/birdcall thresholds for each bird. We used a grid search method to find the best nocall quantile-based threshold of each bird, such that we achieve the maximum AUC score per each bird, or in other words, such that we can distinguish better birdcalls from noise or background, independent from the bird species present in the audio. Figure \ref{figure4} shows the distribution of probabilities using train\_soundscapes. This Class-wise (CW) post-processing method further boosted our score in comparison to Penalization and is more robust.

\section{Results and discussion}
Table \ref{tab:experiments} summarizes our experiments. We tested 9 different CNN backbones. We found difficult to calibrate thresholds using an ensemble of models, yet, we used quantile-based thresholds on the ensemble predictions. Penalization (PN) showed good performance in Public LB. However, penalizing common birds as \textit{skylar} or \textit{houfin} that most probably appear in most of the audios is not realistic. On the other hand, the Class-wise (CW) method showed better performance in general, and it is robust to background birds. We find that calibration of thresholds is very sensitive because there are very few rare birds in audio clips and the real world. Our results imply that we can find proper thresholds for each rare bird using nocall/birdcall validation and a quantile-based approach without strongly labeled data, as we show in Figure~\ref{figure4}.

\begin{figure}[!ht]
    \centering
    \setlength{\tabcolsep}{2.0pt}
    \begin{tabular}{c}
    \includegraphics[width=\linewidth]{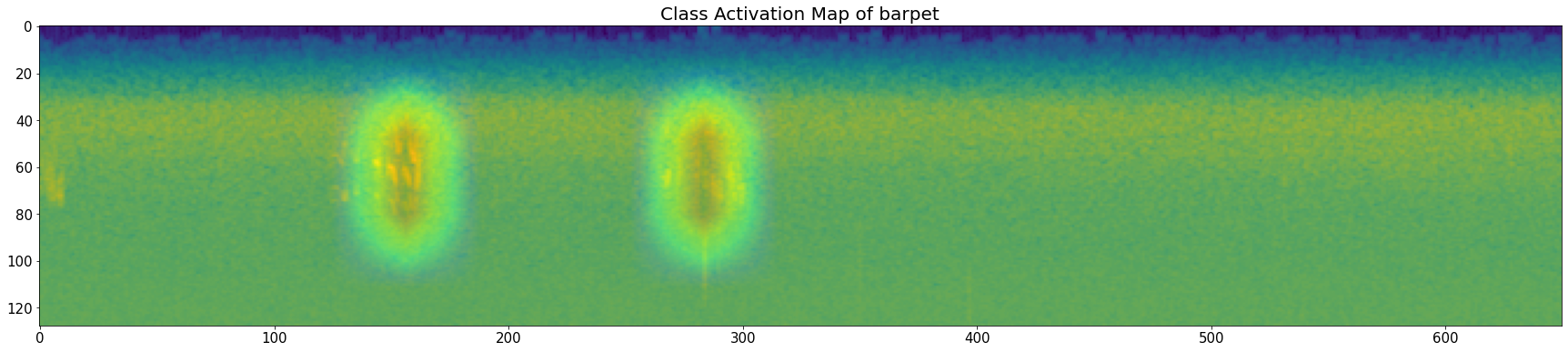} \tabularnewline
    \includegraphics[width=\linewidth]{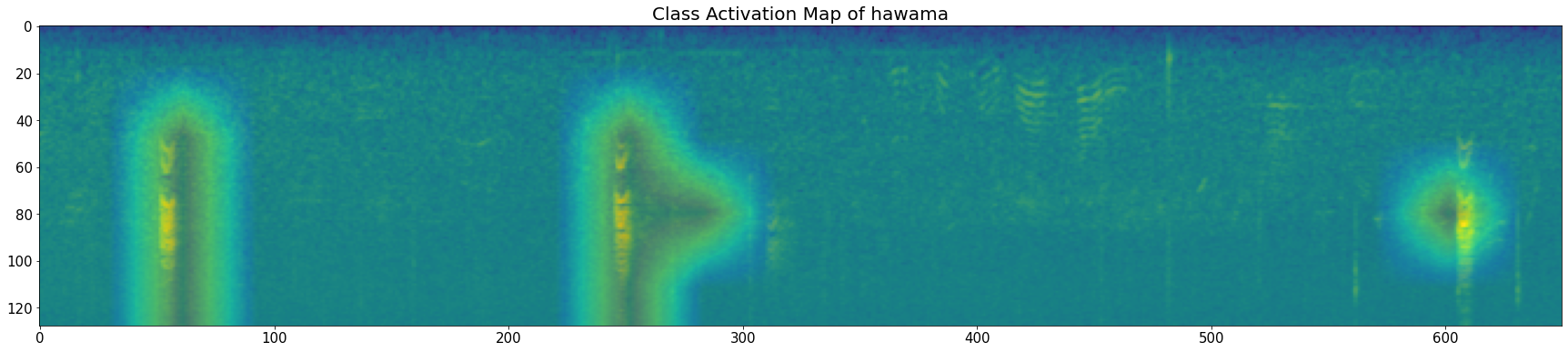} \tabularnewline
    \includegraphics[width=\linewidth]{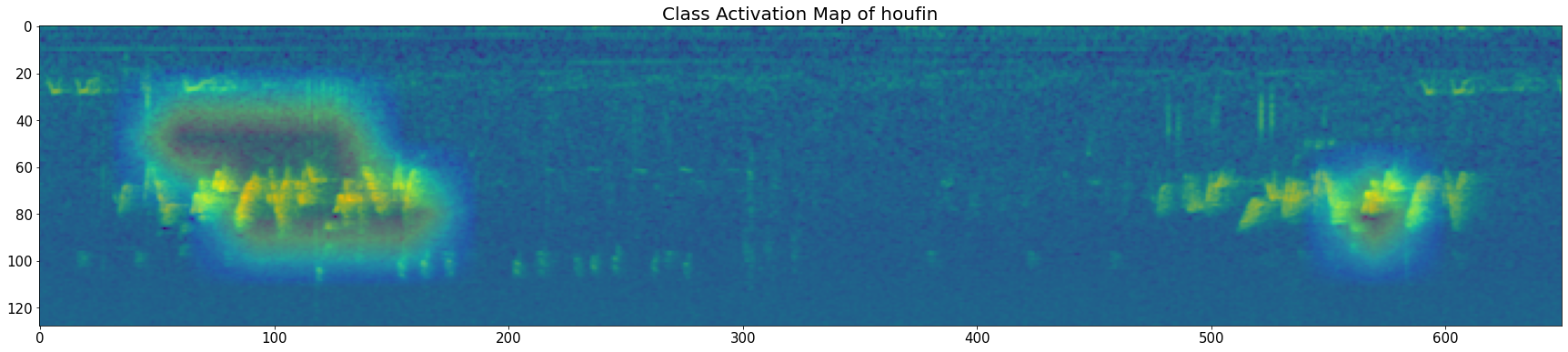} \tabularnewline
    \includegraphics[width=\linewidth]{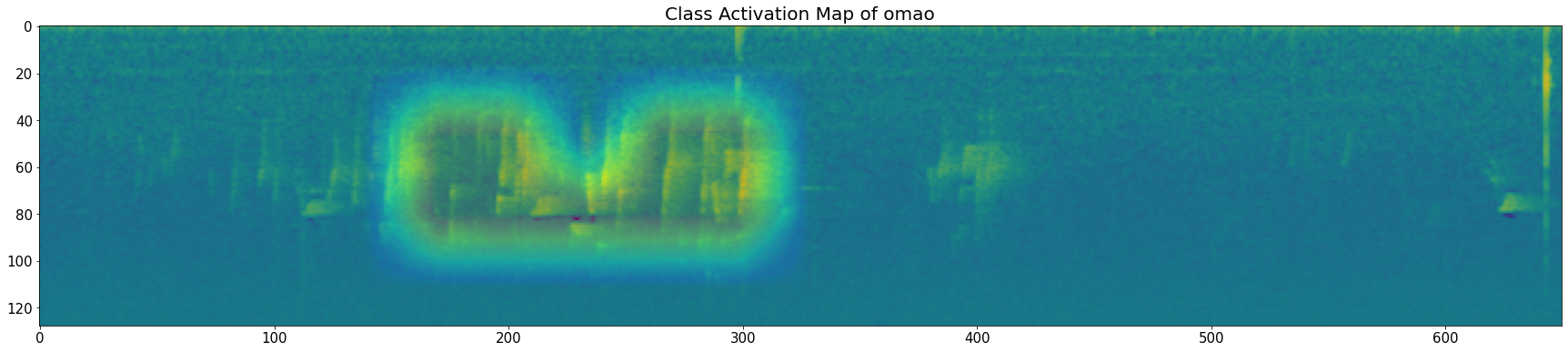} \tabularnewline
    \includegraphics[width=\linewidth]{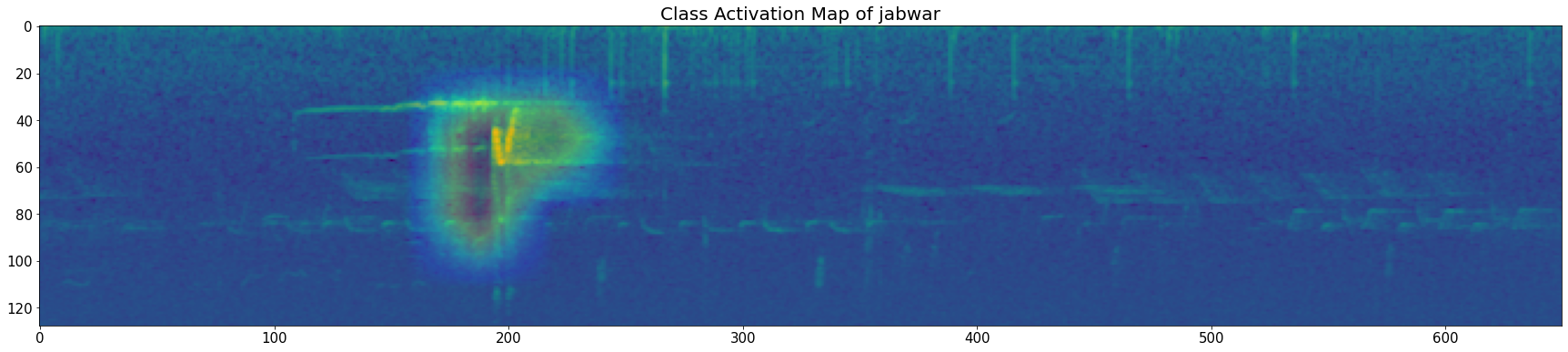} \tabularnewline
    \end{tabular}
    \caption{Grad-CAM~\cite{selvaraju2017grad} activations from our model \texttt{tf\_efficientnet\_b0\_ns} on different validation audio spectrograms. These qualitative results show how our model focuses on particular frequencies through time to recognize the birds. Best viewed in electronic version.}
    \label{fig:gradcams}
\end{figure}

We also provide qualitative Grad-CAM~\cite{selvaraju2017grad} results of our model \texttt{tf\_efficientnet\_b0\_ns} in Figure~\ref{fig:gradcams}, which shows how our model is able to learn and focus on particular frequencies and segments within the audio, and it is robust against background noise.

\begin{table}[!hb]
  \centering
  \caption{Experiments result of models. For local validation, we used "micro F1-score" and train soundscapes. We highlight in {\color{blue}blue} our top-3 models, in {\color{orange}yellow} the results of our final submission ensemble, and in {\color{green}green} the top solutions in the challenge LB. We also distinguish two post-processing methods: PN and CW.
  Contemporary results from other competitors can be found at~\cite{birdCLEF2022krishnan,birdCLEF2022martynov,birdCLEF2022miyaguchi,birdCLEF2022sampathkumar}.}
  
  \begin{tabular}{ |c|c|c|c|c|c| } 
   \hline
   \rowcolor{mygrey} Backbone & CV & Public LB & Private LB & Post-Proc. \\ 
   \rowcolor{LightCyan} \textbf{tf\_efficientnet\_b0\_ns}~\cite{tan2020efficientnet} & 0.8745 & 0.7922 & 0.7240  & PN\\ 
   \rowcolor{LightCyan} \textbf{tf\_efficientnet\_b0\_ns}~\cite{tan2020efficientnet} & 0.8745 & 0.7817 & 0.7548  & CW\\ 
   \rowcolor{LightCyan} eca\_nfnet\_l0 \cite{nfnet} & 0.8761  & 0.7510 & 0.7387 & CW\\
   \rowcolor{LightCyan} resnest50d~\cite{zhang2020resnest} & 0.8822& 0.7550 & 0.7372 & CW\\ 
   tf\_efficientnet\_b1\_ns~\cite{tan2020efficientnet} & 0.7843 & 0.7395 & 0.6946 & CW\\ 
   tf\_efficientnet\_b2\_ns~\cite{tan2020efficientnet} & \textbf{0.8848} & 0.7277 & 0.7046 & CW\\ 
   tf\_efficientnet\_b3\_ns~\cite{tan2020efficientnet} & 0.8561 & 0.7262 & 0.6640 & CW\\ 
   tf\_efficientnetv2\_s\_in21k~\cite{tan2021efficientnetv2} & 0.8632 & 0.7620 & 0.7439 & CW \\ 
   tf\_efficientnetv2\_b0~\cite{tan2021efficientnetv2} & 0.8762 & 0.7268 & 0.7070 & CW\\ 
   \rowcolor{LightYellow} Ours Ensemble & - & 0.7971 & \textbf{0.7733} & CW\\ 
   \rowcolor{LightYellow} Ours Ensemble & - & \textbf{0.8359} & 0.7630 & PN\\ 
   \rowcolor{mygreen} 1st Place & - & 0.8953 & 0.8527 & \\ 
   \rowcolor{mygreen} 2nd Place & - & 0.9128 & 0.8438 & \\
   \rowcolor{mygreen} 3rd Place & - & 0.8750 & 0.8126 & \\
   \rowcolor{mygreen} BirdNet~\cite{birdnet} & - & 0.85 & 0.78 & \\
   \hline
\end{tabular}
\label{tab:experiments}
\end{table}

\section{Conclusion}

We hope our work can help researchers and conservation practitioners accurately survey population trends, so they can regularly and more effectively evaluate threats. We present a sound detection and classification pipeline for analyzing soundscape recordings. Our models learn from few data and weak labels; they can accurately classify fine-grained bird vocalizations in 0.04s using a single GPU. Moreover, they show robustness against noisy sounds (e.g., rain, cars). We aim to improve the model's efficiency for smartphone devices applications.

\clearpage

\begin{acknowledgments}

Marcos Conde is supported by H2O.ai and by Humboldt Foundation (JMU Würzburg).\\
We would like to thank Kaggle and Dr.~Stefan Kahl for hosting the BirdCLEF 2022 Challenge. We also want to thank the contributions from: Amanda K. Navine, Ann Tanimoto-Johnson, Hidehisa Arai, Christof Henkel, Pascal Pfeiffer, and Philipp Singer.
\end{acknowledgments}

\bibliography{references}

\end{document}